# Emergent Cooperative Superstructures via Order–Disorder Kinetics in Molecule-Intercalated NbSe$_2$

Taiga Ueda[1,2], Hideki Matsuoka[1,†], Shungo Aoyagi[1,2], Shunsuke Kitou[3], Yijin Zhang[1,4], Fumihiko Kimura[2], Kenta Hagiwara[2], Masato Sakano[5], Takahiro Iwagaki[1,2], Yuiga Nakamura[6], Kyoko Ishizaka[2,7], Tomoki Machida[1], Masayuki Suda[8], Taka-hisa Arima[3,7] and Naoya Kanazawa[1]

[1] *Institute of Industrial Science, The University of Tokyo, Tokyo 153-8505, Japan*
[2] *Department of Applied Physics, The University of Tokyo, Tokyo 113-8656, Japan*
[3] *Department of Advanced Materials Science, The University of Tokyo, Kashiwa, 277-8561, Japan*
[4] *Department of Physics, The University of Tokyo, Tokyo 113-0033, Japan*
[5] *Graduate School of Informatics and Engineering, The University of Electro-Communications, Tokyo 182-8585, Japan*
[6] *Japan Synchrotron Radiation Research Institute (JASRI), SPring-8, Hyogo 679-5198, Japan*
[7] *RIKEN Center for Emergent Matter Science, Wako 351-0198, Japan*
[8] *Department of Chemistry, Graduate School of Science, Nagoya University, Nagoya 464-8602, Japan*

† To whom correspondence should be addressed. E-mail: hideki-m@iis.u-tokyo.ac.jp

**Abstract**

The design of quantum states at heterointerfaces has enabled a variety of emergent phenomena. Among them, molecular intercalation superlattices[1–3] have attracted attention as tunable hybrid materials, formed by inserting organic molecules into van der Waals crystals, where molecular structure and chemistry provide new degrees of freedom. Traditionally, the intercalated molecules have been regarded as inactive spacers, while possible molecular ordering and its impact on the host lattice have remained largely unexplored. Here, we report the discovery of a cooperative superstructure (CSS) phase in molecule intercalated $NbSe_2$, where ordering of the guest molecules induce a concomitant superstructure in the $NbSe_2$ host lattice, characterized by a moiré structure due to incommensurability between the molecular layer and the inorganic lattice. Synchrotron X-ray diffraction reveals the emergence of CSS phase, accompanied by crystal symmetry lowering. Complementary resistivity and thermal-quench measurements show that the transition is governed by unusually slow order-disorder kinetics, so that the CSS phase can be selectively accessed under standard laboratory cooling rates. This kinetic behavior arises from slow molecular dynamics coupled to the host lattice, contrasting with fast charge or magnetic ordering in inorganic solids[4]. Our findings establish molecular ordering as a route for engineering heterointerfaces, enabling thermally programmable superstructures.

The control of quantum states through interface engineering in heterostructures and superlattices has opened pathways to a wide variety of emergent phenomena. Prominent examples include two-dimensional electron gases at oxide interfaces[5,6], proximitized interfaces with magnetism, superconductivity, or topological surface states[7–9], and moiré superlattices created in van der Waals (vdW) heterostructures[10–12]. A defining feature of such heterointerfaces is the emergence of properties entirely distinct from those of the constituent materials, often yielding functionalities beyond the simple sum of their parts.

Among various engineered heterointerfaces, organic-inorganic hybrids have emerged as a distinctive family, especially in the form of molecular intercalation superlattices (MISs) realized by inserting organic molecules into the vdW gaps of layered crystals[1–3,13–27]. Conventionally, MISs have been viewed as a route to create *bulk monolayer materials* which can exhibit effectively two-dimensional properties even in bulk forms because the inorganic layers are spatially separated by the intercalated molecules. Illustrative cases are the emergence of excitonic photoluminescence in $MoS_2$[25], Ising superconductivity in $NbSe_2$[21], and band-gap modulation in black phosphorus[16]. More recently, MISs have enabled symmetry engineering. For instance, the insertion of chiral molecules has been reported to induce chiral-induced spin selectivity (CISS)[23,26] and even unconventional superconductivity[27], broadening the design space for quantum functionalities and phases.

While previous studies demonstrated that molecular intercalation can modulate dimensionality or embed molecular asymmetry into the host lattice, such approaches primarily focused on the static influence of individual molecular species. In contrast, how molecular ensembles collectively organize and dynamically interact with the host lattice has yet to be clarified. Beyond individual-molecule effects, collective molecular behavior, as commonly observed in soft-matter systems, represents an uncharted territory in MISs. When confined within the vdW gaps, molecules can form distinct collective states, ranging from disordered liquid- or glass-like arrangements to ordered crystalline phases. Yet, the nature of phase transition among these possible collective states and their coupling to the host lattice, remain open questions central to understanding hybrid quantum materials.

In this study, we identify a novel phase transition in molecular-ion-intercalated $NbSe_2$ that yields a dual-scale cooperative superstructure (CSS), resolved by synchrotron X-ray diffraction. The CSS comprises two coupled modulations: molecular ordering, namely an equidistant arrangement of molecules, and a long-period distortion of the $NbSe_2$ host-lattice. These modulations emerge simultaneously, and the long-period lattice distortion constitutes an intrinsic moiré superstructure that originates from the incommensurability between the molecular ordering layer and the inorganic lattice. Temperature-dependent resistivity measurements, including controlled thermal-quench protocols, reveal slow order-disorder kinetics, enabling selective stabilization of either the equilibrium CSS or a supercooled metastable disordered state by varying the cooling rate.

**Discovery of the Cooperative Superstructure (CSS) Phase**

The target system is (*S*)-MBMIm/NbSe$_2$, in which the chiral organic cation (*S*)-1-(2-Methylbutyl)-3-methylimidazolium [(*S*)-MBMIm] is electrochemically intercalated into the vdW gaps of 2*H*-NbSe$_2$ (Figs. 1**a** and 1**b**). The intercalation setup is described in detail in Methods. The structural and vibrational properties of (*S*)-MBMIm/NbSe$_2$ were characterized by X-ray diffraction (XRD) and Raman spectroscopy, respectively. As shown in Fig. 1**c**, the intercalation leads to a pronounced shift of all (0,0,*l*) reflections toward lower diffraction angles in the XRD patterns, corresponding to a substantial increase in the interlayer spacing of (*S*)-MBMIm/NbSe$_2$ (11.6 Å) compared with pristine NbSe$_2$ (6.27 Å). No remanent reflections from the pristine NbSe$_2$ are observed after intercalation, indicating that the intercalated phase extends throughout the entire crystal, confirming bulk homogeneity. A similar *c*-axis expansion is also observed for other organic intercalants (Extended Data Fig. 1) and represents a common feature of MISs across vdW layered materials[16–20]. Furthermore, Raman spectra reveal the disappearance of the shear mode and a shift of the A$_{1g}$ and E$_{2g}$ peaks upon intercalation (Fig. 1**d**). These spectral changes closely resemble those of isolated NbSe$_2$ monolayer[28] and have also been observed in NbSe$_2$ intercalated with other organic molecules[21]. These results confirm that (*S*)-MBMIm molecules are uniformly intercalated into NbSe$_2$ lattice, separating the vdW layers and electronically decoupling them as further corroborated by photoemission spectroscopy (Supplementary Note 2).

We observe a structural phase transition upon cooling in (*S*)-MBMIm/NbSe$_2$ by synchrotron XRD (Figs. 1**e**–1**j**). To aid understanding of the complex diffraction patterns that follow, we first summarize our interpretation before presenting the experimental evidence. Here we refer to the low-temperature phase as the CSS phase, which is characterized by two distinct orders developing in concert: the intercalated molecular layer crystallizes, perhaps forming a plastic crystal, from a disordered liquid-like state (molecular ordering, MO); this molecular ordering is accompanied by a periodic lattice distortion in the NbSe$_2$ layers that produces superstructure (SS) reflections (Fig. 1**e**). The two in-plane modulations are mutually incommensurate: a short period of the molecular layers ($a_{MO}$ = 6.29 Å) and a long period of the host lattice distortion ($a_{SS}$ = 32.7 Å). The long period $a_{SS}$ follows naturally from the lattice mismatch between the host NbSe$_2$ ($a$ = 3.46 Å) and the ordered molecular layer ($a_{MO}$), in analogy with other lattice-mismatch moiré superlattices[29,30], as discussed later. Figures 1**f** and 1**g** schematically depict the reciprocal-space maps of the high-temperature disordered phase and the low-temperature CSS phase, respectively, highlighting the appearance of MO and SS reflections relative to the fundamental Bragg peaks of the NbSe$_2$ host.

We next present the evidence supporting this interpretation. Figures 1**h**–1**j** show diffraction patterns of (*S*)-MBMIm/NbSe$_2$ in the low-temperature CSS phase on the *h* 0 *l*, *h k* 13, and *h k* 0 planes at 100 K, respectively. Two distinct sets of additional diffraction peaks are identified, *i.e.*, the MO and SS peaks introduced above. The origin of each peak type can be clearly distinguished from its intensity profiles. The MO peaks appear only around the origin (0,0,0) with sixfold in-plane wavevectors with $|q_{MO}|$ = 0.551 $a^*$ (Figs. 1**h** and 1**j**), corresponding to an in-plane molecular-order period $a_{MO}$ = 6.29 Å (Fig. 1**e**). Their assignment to molecular order is supported by the rapid decay of intensity at large scattering vectors $Q$, which is characteristic of the steeply $Q$-dependence of the atomic form factors of the light elements such as C, N, and H. In addition, the possible orientational and positional disorder inherent in the molecular sublattice is likely to further enhance this steep $Q$-dependence. Furthermore, the diffuse elongation of intensity along the $c^*$-axis (Fig. 1**h**) indicate that the molecular ordering is essentially two-dimensional without interlayer correlation. Taken together, these observations demonstrate that the MO peaks directly reflect the ordering of molecules within the van der Waals gap that has been largely overlooked in previous studies of MISs.

In contrast, the SS peaks form sharp satellites even around (0,0,*l*) reflections at large *l*, also showing sixfold symmetry with $|q_{SS}|$ = 0.106 $a^*$ (Figs. 1**h** and 1**i**). The SS peaks arise from a host lattice distortion with an in-plane period $a_{SS}$ = 32.7 Å (Fig. 1**e**), as confirmed by their persistence to large $Q$, which reflects the less $Q$-dependent form factors of the heavy Nb and Se atoms. We emphasize that the wavevectors of the NbSe$_2$ host lattice, the MO, and the SS peaks in reciprocal space satisfy the approximate relation $|q_{SS}| \simeq 2q_{MO} - a^*$, indicating that the SS corresponds to a moiré pattern produced by the slight periodicity mismatch between the molecular ordering and twice the NbSe$_2$ lattice constant (see Extended Data Fig. 2). Notably, the SS peaks predominantly appear around out-of-plane Bragg peaks such as (0,0,13) (Fig. 1**i**), demonstrating that the associated lattice distortion involves atomic displacements primarily along the $c$-axis. Our diffraction simulations are consistent with sinusoidal-type or breathing-type modulation that can be described by either a single-$q$ vector with domains related by threefold rotation or a triple-$q$ state with single domain, as schematically shown in Fig. 1**e** (also see Supplementary Note 3 and Supplementary Note 4). Similar sinusoidal-type modulations have been reported in other MISs and van der Waals superlattices[30–32], possibly because the molecular asymmetry differentially couples to the upper and lower surfaces of each layer and induces local bending perturbations.

**Phase transition of the CSS and associated symmetry breaking**

To elucidate the structural phase transition and the cooperative coupling between molecular and NbSe$_2$ lattice orders, we analyze the temperature evolution of these orders and the accompanying changes in crystallographic symmetry in ($S$)-MBMIm/NbSe$_2$. Figure 2**a** compares reciprocal-space maps at 100 K and 320 K. At 100 K, both the MO and SS peaks are clearly observed, whereas they disappear completely at 320 K. This difference is more evident in the line profiles (Figs. 2**b** and 2**c**), indicating a well-defined transition between the ordered and disordered phases occurring between 100 K and 320 K. The temperature evolution of the diffraction intensities of the MO and SS peaks during heating (Fig. 2**d**) shows that both sets of reflections vanish simultaneously at the transition temperature $T_m \sim 310$ K, corresponding to the thermal melting of the molecular order. $T_m$ exhibits a sample-dependent variation within the range 310–320 K, as provided in Supplementary Table 1.

As another structural signature of the transition, we identify the breaking of the $c$-glide symmetry in the NbSe$_2$ host lattice. The $c$-glide symmetry corresponds to the combination of a mirror operation and a translation along the $c$-axis, which is preserved in pristine NbSe$_2$ with the space group $P6_3/mmc$[33] (Fig. 2**e**). The presence or absence of this symmetry can be recognized from the extinction rules in the XRD data: when the $c$-glide symmetry is preserved after the intercalation, diffraction peaks at ($h,h,l$) with odd $l$ are forbidden (Fig. 2**f**); whereas breaking of the symmetry leads to the appearance of these reflections (Fig. 2**g**). Figure 2**h** shows reciprocal-space maps around the (1,1,$l$) positions. In the high-temperature phase at 320 K, the extinction rule is satisfied, indicating that the $c$-glide symmetry remains intact. By contrast, in the CSS phase at 100 K, reflections at (1,1,$l$) with odd $l$ become clearly visible. Importantly, the emergence of these formerly forbidden reflections coincides with the appearance of the MO and SS peaks in the temperature change (see Extended Data Fig. 3), demonstrating that the formation of CSS state is accompanied by the breaking of the $c$-glide symmetry.

From the extinction condition, we discuss the crystallographic symmetry of ($S$)-MBMIm/NbSe$_2$. The pristine NbSe$_2$ crystallizes in the centrosymmetric space group $P6_3/mmc$[33], and the high-temperature phase of ($S$)-MBMIm/NbSe$_2$ retains the same space group (see Supplementary Note 5). On the other hand, the breaking of the *c*-glide symmetry in CSS phase results in reduced symmetry: $P6_322$, $P6_3/m$ or lower (see Supplementary Note 6). Such reduction of symmetry can originate either from SS formation within the NbSe₂ host itself, as in the charge-density-wave transition of 1*T*-TaS₂[34], or from molecular ordering within the intercalated layer, as seen in alkali-ion–intercalated graphite[35]. While our XRD data establishes the key symmetry elements, determining the precise atomic displacements will require local probes such as transmission electron microscopy (TEM) or scanning tunneling microscopy (STM).

**Kinetics of CSS phase transition**

The cooperative phase in (*S*)-MBMIm/NbSe₂ is governed by slow molecular ordering kinetics, which strongly influences the cooling-rate dependence of CSS formation. To investigate this kinetic effect, we perform thermal quench experiments. In a first-order transition, the transformation involves overcoming an activation barrier between two competing phases. Because the nucleation and growth proceed at a finite speed, the transformation timescale can compete with the externally imposed cooling rate. When cooling is too rapid, the transition cannot complete, and the high-temperature phase becomes trapped as a quenched metastable state[36–38] (Fig. 3**a**). Suppression of first-order phase transitions by rapid cooling is commonly observed in systems where the kinetics are relatively slow, such as order-disorder transitions in polymeric soft matter[39,40] and charge-order transitions in some organic conductors[4,41,42]. In contrast, electron or spin phase transitions in inorganic crystalline solids generally evolve too fast to be kinetically hindered within the standard experimental cooling-rate range ($2 \times 10^{-3}$ to $4 \times 10^{-1}$ K s⁻¹). In such fast systems, suppression of the transition is achieved through extremely rapid cooling methods, including ultrafast cooling[43–45] or sample miniaturization to suppress nucleation of the competing phase[46–48].

Thermal quench effects on the CSS transition in (*S*)-MBMIm/NbSe₂ were evaluated using both synchrotron XRD and temperature-dependent resistivity measurements (see Extended Data Fig. 4 for the correspondence between diffraction and resistivity signatures of the CSS transition). Figure 3**b** shows the resistivity as a function of temperature under two different cooling rates: slower cooling at 2 K min⁻¹ (~ 0.03 K s⁻¹) and faster cooling at 11-20 K min⁻¹ (~ 0.18-0.33 K s⁻¹). Under the slower cooling condition, a pronounced anomaly corresponding to the CSS transition is observed: the resistivity decreases continuously over the range *T* = 320–260 K with a particularly sharp drop around 265 K. By contrast, under the faster cooling, the anomaly is strongly suppressed, indicating that the CSS transition is avoided. Synchrotron XRD corroborates this kinetic suppression. As shown in the inset of Fig. 3**b**, under a rapid cooling condition (20 K min$^{-1}$), neither the MO peaks nor the SS peaks are observed, consistent with the resistivity data and demonstrating that the CSS transition is inhibited once the cooling rate exceeds a critical threshold (see Extended Data Fig. 5 for the detailed cooling rate dependence).

To clarify the kinetics underlying this cooling-rate dependence, we examine the time evolution of CSS formation from a supercooled disordered state. To prepare the supercooled state, the sample was first heated to 400 K to erase the CSS state, then cooled to a target temperature with 11 K min$^{-1}$ and held at that temperature while recording the resistivity as a function of time. This procedure was repeated for a series of target temperatures. Figure 4**a** displays the time evolution of resistivity, where the magnitude of resistivity change $-\Delta\rho_{xx}/\rho_{xx}(t=0)$ tracks the progress of CSS formation. At sufficiently low ($T$ = 220 K) or high ($T$ = 320 K) temperatures, the resistivity shows negligible temporal variation. In contrast, pronounced time-dependent changes are observed in the intermediate range of 260–290 K, where the CSS transition proceeds. Based on these measurements, we construct a time-temperature-transformation (TTT) diagram (Fig. 4**b**), which identifies the time-temperature domain where the transformation occurs.

A noteworthy feature is that the CSS transition is suppressed even at a relatively modest cooling rate of 20 K min$^{-1}$, which falls within the range of standard laboratory protocols. As shown in the TTT diagram (Fig. 4**b**), the trajectory corresponding to 20 K min$^{-1}$ does not intersect the time-temperature domain of CSS formation; by contrast, the 2 K min$^{-1}$ trajectory crosses this domain, providing sufficient time for the transition to proceed. This indicates that the kinetics of the CSS transition are significantly slower than those of conventional phase transitions in inorganic crystalline solids. To place this observation in a broader context, Fig. 4**c** summarizes the cooling rates required to suppress various phase transitions, ranging from organic molecular systems to crystalline solids. In particular, to our knowledge, suppression of charge-density-wave (CDW) or structural transitions in layered vdW materials has not been reported within standard laboratory cooling-rate ranges. Only one notable exception is the commensurate CDW transition in 1$T$-TaS$_2$, which can be thermally suppressed, but only under extremely rapid cooling conditions in the bulk form, typically exceeding 50 K min$^{-1}$ (Refs. 44,49). We attribute the unusually slow kinetics of the CSS transition primarily to the intrinsically slow molecular ordering within the vdW gap. This interpretation aligns with the well-known tendency of molecular and polymeric systems to have slow crystallization, which makes their phase transitions readily inhibited by rapid cooling relative to inorganic materials. These slow molecular kinetics constitute the rate-determining factor and govern the overall transformation process through coupling in the MIS architecture, distinguishing the CSS transition from purely electronic or structural transitions in inorganic hosts.

**Conclusion and Outlook**

By resolving ($S$)-MBMIm/NbSe$_2$ by synchrotron XRD, we identify a dual-scale CSS in which collective molecular ordering of the intercalated layer and a concomitant superstructure of the NbSe$_2$ host co-emerge and drive a first-order phase transition. The wavevector relationship confirms that the long-period superstructure is a moiré pattern arising from the incommensurability between the molecular layer and the NbSe$_2$ lattice. show that their intensities vanish simultaneously at the order-disorder temperature $T_m$ = 310-320 K. Thermal quenching protocols establish that slow molecular ordering kinetics set the transition timescale, so the equilibrium CSS state and a supercooled disordered state can be repeatedly switched using standard laboratory cooling rates.

Beyond this specific chemistry, the mechanism is likely generic. Diffraction studies have rarely been used to search for collective molecular ordering in MISs, yet we observe analogous ordering in a related imidazolium intercalant, BMIM (Supplementary Note 7). Systematically varying the intercalant should reveal which molecular species cause cooperative ordering and how the timescale depends on molecular parameters such as length, symmetry, and electronegativity; tuning these parameters is expected to modulate the accessible cooling-rate window. The MISs will realize thermally programmable control of heterointerface states and offer a route to stabilize various hidden metastable phases in host inorganic materials.

More broadly, embedding molecular degrees of freedom as active kinetic elements in vdW hosts offers a route to engineer symmetry breaking, transition dynamics, and interface electronic states on demand. The CSS concept therefore bridges soft-matter ordering and quantum materials, and enables the rational design of hybrid interfaces whose emergent phases and switching functionalities can be actuated by temperature through molecule-lattice coupling.

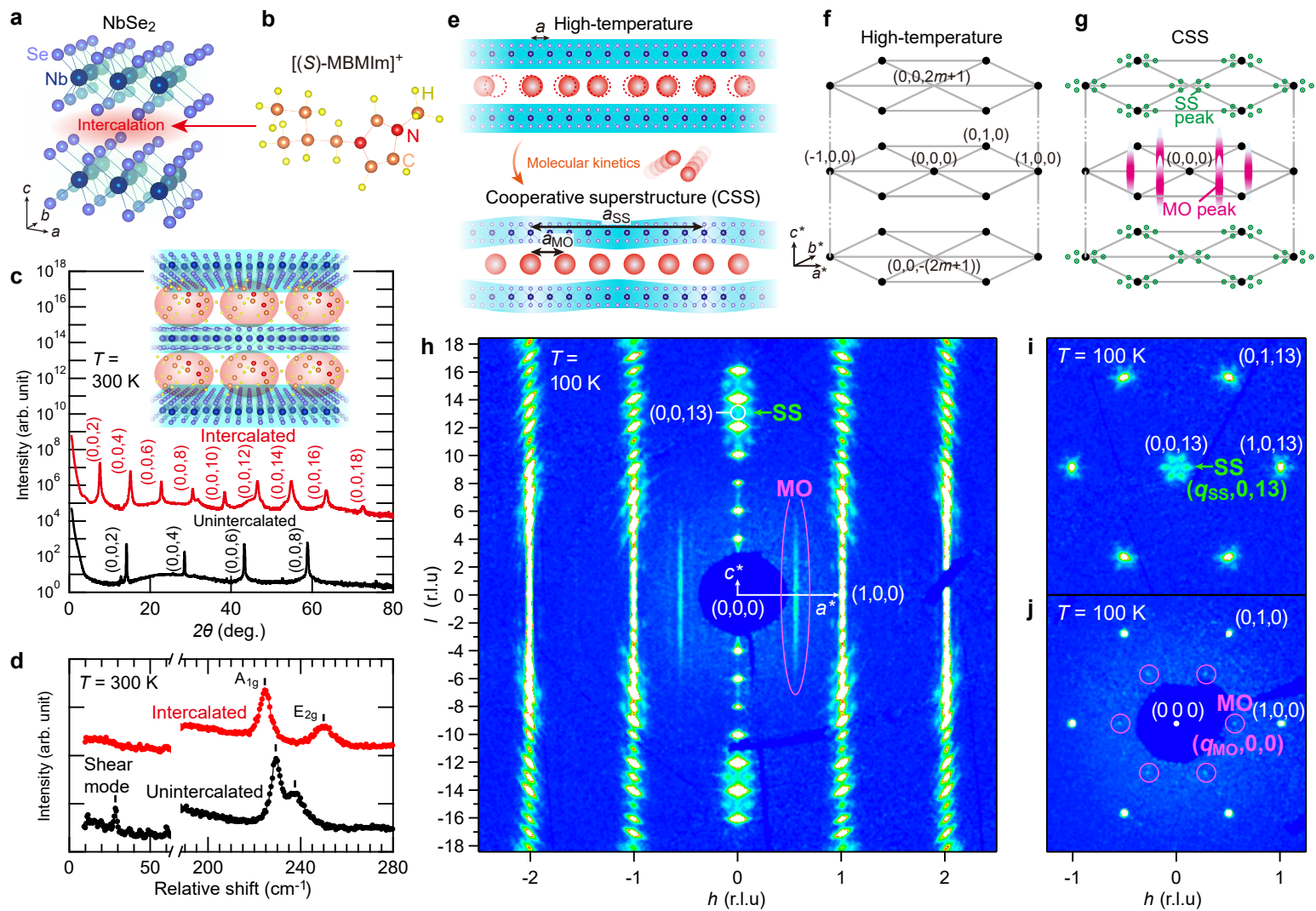

**Fig. 1 | Structural and vibrational characterization of (*S*)-MBMIm/NbSe$_2$.**
**a, b**, Crystal sturcture of NbSe$_2$ (**a**) and molecular structure of the chiral organic cation (*S*)-MBMIm (**b**) used for intercalation. The crystal structure is drawn by VESTA[50]. **c,** X-ray diffraction (XRD) patterns of pristine and intercalated NbSe$_2$ along the 0 0 *l* direction at 300 K. **d,** Raman spectra of pristine and intercalated NbSe$_2$ at 300 K, highlighting the disappearance of the shear mode and shifts of the A$_{1g}$ and E$_{2g}$ peaks. **e,** Schematic illustration of the cooperative superstructure (CSS), associated with molecular ordering (MO) and superstructures (SS) in host materials. **f, g,** Schematic reciprocal-space diffraction patterns of (*S*)-MBMIm/NbSe$_2$ in the high-temperature phase (**f**) and CSS phase (**g**). The MO peaks (magenta) arise from the ordering of intercalated molecules, and the SS peaks (green) originate from the host-lattice superstructure. **h,** Synchrotron XRD pattern in the *h* 0 *l* plane at 100 K, showing both MO and SS peaks. **i, j,** The *h k* 13 plane highlighting SS peaks (**i**) and the *h k* 0 plane highlighting MO peaks (**j**).

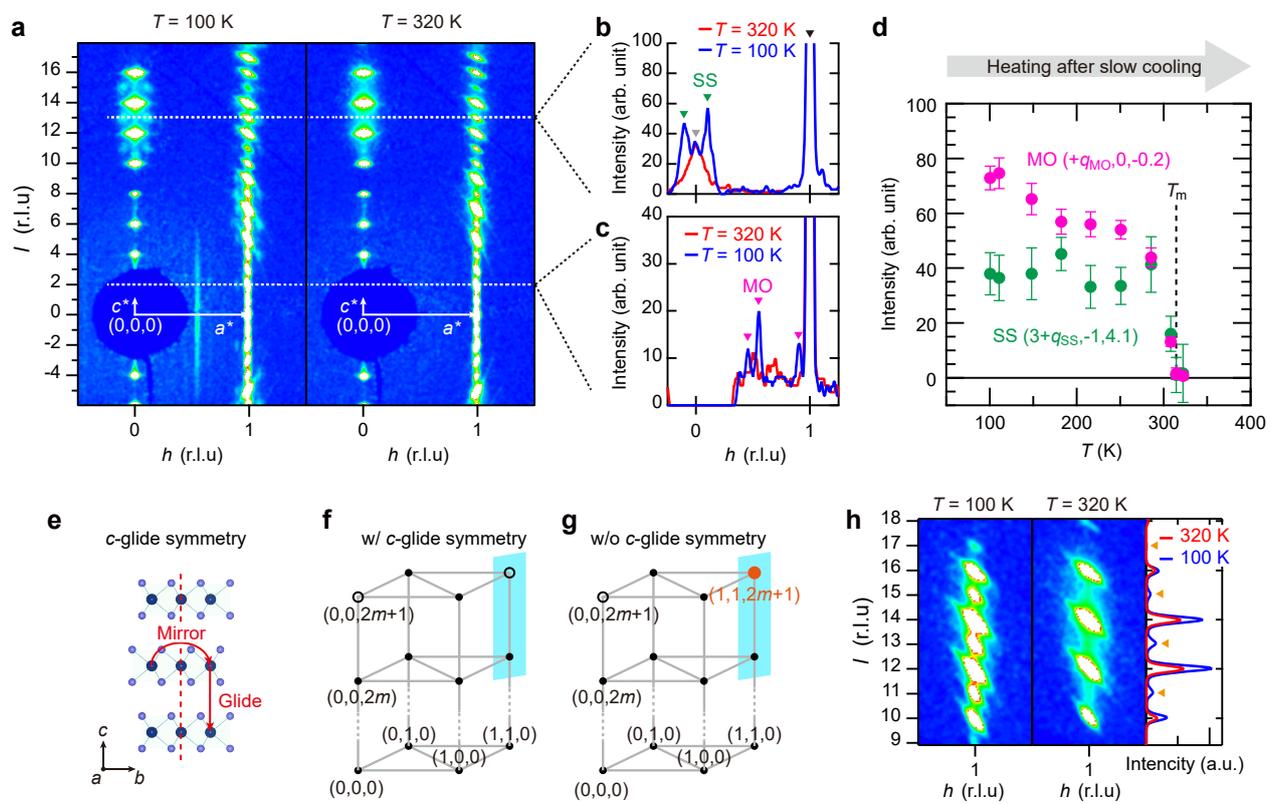

**Fig. 2 | Temperature-dependent phase transition and symmetry breaking of CSS.**
**a,** Synchrotron XRD patterns of (*S*)-MBMIm/NbSe$_2$ measured in the $h\ 0\ l$ plane at 100 K and 320 K. **b, c,** Line profiles along the $h\ 0\ 13$ and $h\ 0\ 2$ directions, highlighting SS peaks (green triangles, **b**) and MO peaks (magenta triangles, **c**), respectively. A black triangle marks a fundamental Bragg reflection of NbSe$_2$, while a gray triangle marks a symmetry-forbidden reflection that appears that arises from the intensity tails of the strong (0 0 12) and (0 0 14) reflections. **d,** Temperature dependence of the diffraction intensities of the MO and SS peaks during heating after slow cooling. Both sets of peaks disappear simultaneously at the melting temperature $T_\mathrm{m} \sim 310$ K. **e,** The schematics of *c*-glide symmetry in NbSe$_2$. **f, g,** Schematic illustration of the extinction rule of XRD associated with *c*-glide symmetry. **h,** Synchrotron XRD reciprocal-space maps in the $h\ h\ l$ plane around the (1,1,*l*) positions at 100 K and 320 K.

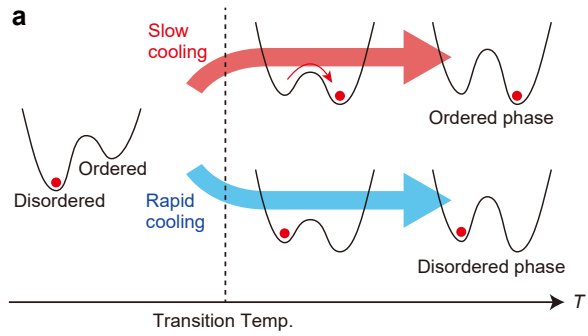

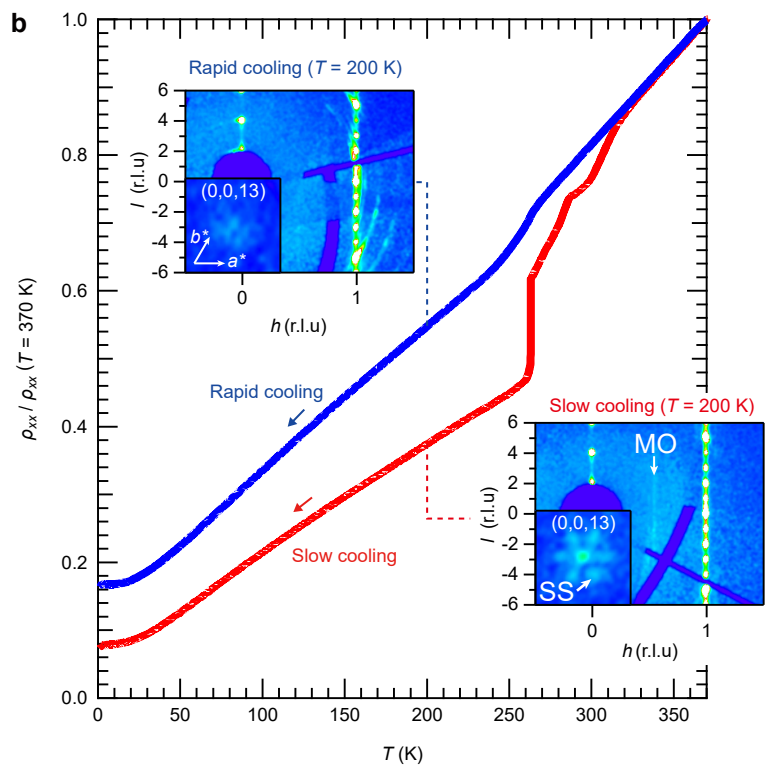

**Fig. 3 | Suppression of the cooperative superstructure (CSS) transition by rapid cooling.**
**a,** Schematic illustration of the effect of cooling rate on the first-order phase transition. **b,** Temperature dependence of the normalized resistivity of (*S*)-MBMIm/NbSe$_2$ under slow cooling (2 K min$^{-1}$, red) and rapid cooling (11 K min$^{-1}$, blue). Insets show synchrotron XRD maps in the *h* 0 *l* plane at 200 K for samples cooled at slow cooling rate (2 K min$^{-1}$, red) and rapid cooling rate (20 K min$^{-1}$, blue).

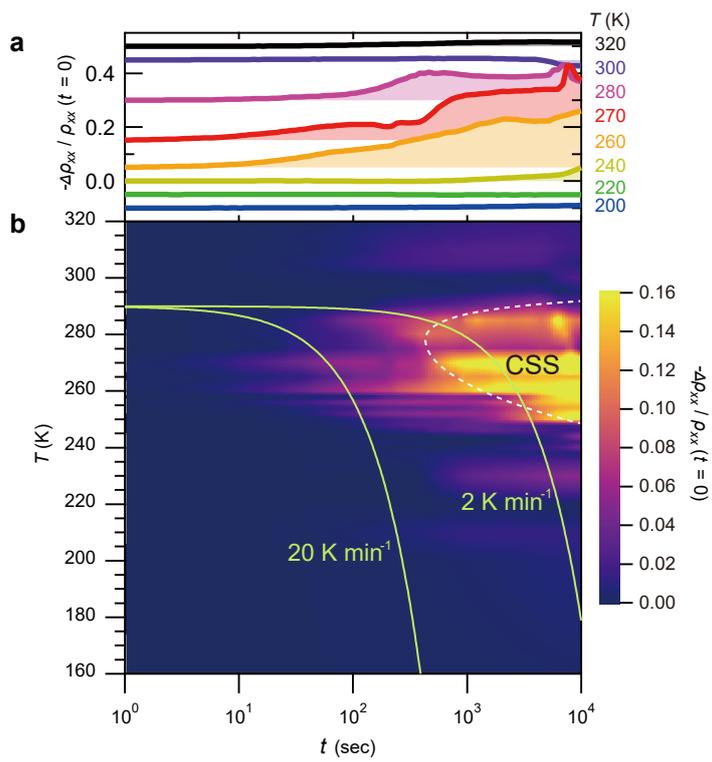
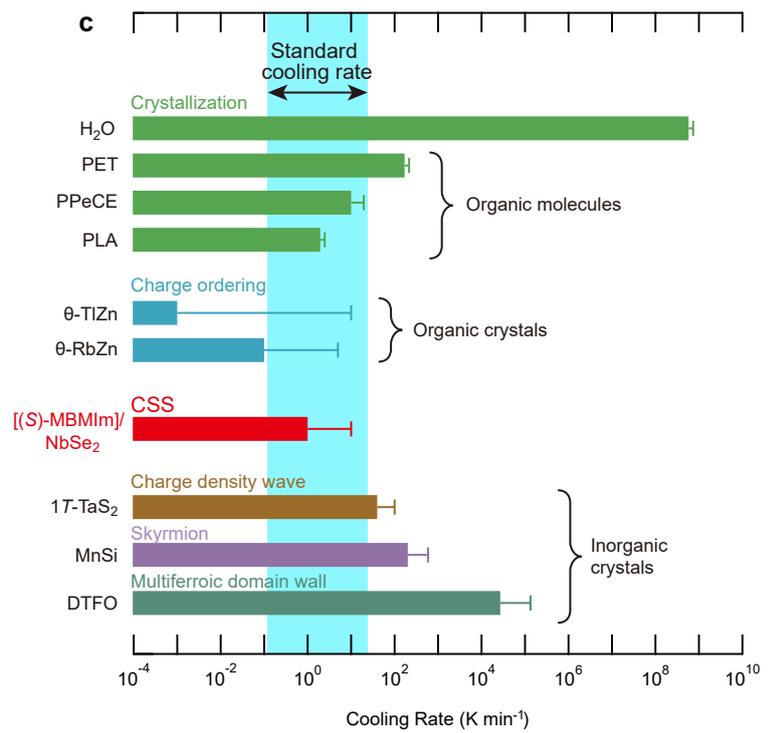

**Fig. 4 | Time-temperature-transformation (TTT) diagram and enumeration of cooling rate dependent phase transitions.**
**a,** Time-dependent resistivity measurements of (*S*)-MBMIm/NbSe$_2$ at fixed temperatures obtained after quenching with rapid cooling from 400 K. **b,** Time-temperature-transformation (TTT) diagram constructed from the measurements of the time-evolution of resistivity. The CSS transformation proceeds in the intermediate range of 260–290 K (dashed line). **c,** Comparison of cooling-rate thresholds for suppressing phase transitions across different classes of materials (reproduced from Refs. 40–43,45,49). Because the regions between the slow-cooling and quenched phases is a crossover[36], the boundaries are depicted with error bars. θ-TlZn and RbZn represent organic conductors θ-(BEDT-TTF)$_2$TlZn(SCN)$_4$ and θ-(BEDT-TTF)$_2$RbZn(SCN)$_4$, respectively.

## Methods

### Sample preparation of NbSe$_2$ crystal

High-quality NbSe$_2$ single crystals were grown from 99.99% pure Nb and Se powders by the I$_2$ chemical vapor transport (CVT) in a temperature gradient of 730–700 °C in a sealed quartz tube for 7 days. To ensure stoichiometry, a slight excess of Se (typically 0.2%) was added to the starting materials.

### Synthetic procedures of (*S*)-1-Bromo-2-methylbutane

(*S*)-2-Methyl-1-butanol (70.5 g, 0.800 mol) was placed in a 500 mL round-bottom flask under a nitrogen atmosphere. Phosphorus tribromide (196 g, 0.720 mol, 0.9 equiv.) was added dropwise at −20 °C. The reaction mixture was stirred at 0 °C for 1 h, warmed to room temperature for 2 h, and then heated at 100 °C for 2 h. After cooling to 0 °C, the mixture was quenched with ice and neutralized with saturated NaHCO$_3$. The organic layer was extracted with diethyl ether, washed with water and brine, dried over Na$_2$SO$_4$, filtered, and concentrated to yield (*S*)-1-bromo-2-methylbutane as a clear, colorless liquid (97.9 g, 81%).

### Synthetic procedures of (*S*)-1-(2-Methylbutyl)-3-methylimidazolium-Bromide ([(*S*)-MBMIm]$^+$ [Br]$^-$)

1-Methylimidazole (59.1 g, 1.0 equiv.) and (*S*)-1-bromo-2-methylbutane (97.9 g) were stirred in a 500 mL round-bottom flask at 60 °C for 72 h. The resulting mixture was washed with diethyl ether to afford [(*S*)-MBMIm]$^+$ [Br]$^-$ as a light-yellow viscous liquid (78%).

### Synthetic procedures of (*S*)-1-(2-Methylbutyl)-3-methylimidazolium-Bis(trifluoromethanesulfonyl)imide] ([(*S*)-MBMIm]$^+$ [TFSI]$^-$)

[(*S*)-MBMIm]$^+$ [Br]$^-$ was dissolved in deionized water and stirred at room temperature. An equimolar amount of lithium bis(trifluoromethanesulfonyl)imide (LiTFSI) dissolved in water was added dropwise to the solution. The reaction mixture was stirred for several hours at room temperature. The resulting TFSI$^-$ salt was extracted with dichloromethane, washed with water to remove residual LiBr, dried over Na$_2$SO$_4$, filtered, and concentrated under reduced pressure to afford [(*S*)-MBMIm]$^+$ [TFSI]$^-$ as a viscous liquid.

### Sample preparation of electrochemical intercalation

The electrochemical intercalation of ionic liquid cations was performed in a custom-built two-electrode cell. Platinum sheets were used for both the working electrode and the counter electrode. NbSe$_2$ crystals were attached to the working electrode, which is immersed in the ionic liquid electrolyte. For the intercalation of [(*S*)-MBMIm]$^+$, a constant voltage of -1.2 V was applied to the [(*S*)-MBMIm]-[TFSI] ionic liquid at 100 °C conditions in an ambient atmosphere. For the intercalation of 1-Butyl-3-methylimidazolium ([BMIm]$^+$) and 1-Ethyl-3-methylimidazolium ([EMIm]$^+$), a voltage of -3.5 V was applied to the [BMIm]-[TFSI] (Tokyo Chemical Industry Co., Ltd) and [EMIm]-[TFSI] (Tokyo Chemical Industry Co., Ltd) under the same conditions as [(*S*)-MBMIm]$^+$. To ensure uniform intercalation throughout the bulk crystal, the process was maintained for 6 to 12 hours. Further experimental details are provided in the Supplementary Note 1.

**XRD experiments**

In-house XRD patterns were collected on a RIGAKU SmartLab diffractometer with Cu-K$\alpha$ radiation ($\lambda$ = 1.5418 Å) for determining the interlayer spacings. Synchrotron XRD experiments were conducted at the BL02B1 beamline of SPring-8[51], Japan. A nitrogen gas blowing device was used to cool the crystal to 100 K. Diffraction patterns were recorded using a CdTe PILATUS two-dimensional detector (dynamic range ~10$^6$) with an X-ray energy of $E$ = 35 keV ($\lambda$ = 0.35288 Å) for Figs. 1 and 2, and $E$ = 25 keV ($\lambda$ = 0.49671 Å) for Fig. 3. The intensities of Bragg reflections corresponding to *d*-spacings is 0.46 Å for Figs. 1-2 or 0.44 Å for Fig. 3, which were collected using the CrysAlisPro software[52]. To mitigate radiation damage of the samples, the incident X-ray beam was shuttered except during active data collection. Intensities of equivalent reflections were averaged, and the structural parameters were refined with JANA2006[53].

**Transport measurements**

The electrical resistance was measured using a Quantum Design Physical Property Measurement System (PPMS) at temperatures of 2–400 K under magnetic fields up to 9 T, and a pressure of 1–10 torr. Measurements were performed on bulk samples, with electrical contacts for a Hall bar configuration defined by gold wires fixed by Ag epoxy at distances of several millimeters. The excitation current was applied in the range of 2.5-5 mA, depending on the resistance value. Before each measurement, the sample temperature was raised to 400 K and held at that temperature for 30 minutes to remove contaminants such as water.

To obtain the time-temperature-transformation (TTT) diagram, the sample was first heated to and held at 400 K for 30 minutes, followed by rapid cooling at a rate of 20K min$^{-1}$ to a target temperature, where the time evolution of the resistivity was recorded.

**First-principles calculations**

The first-principles calculations were performed using the WIEN2k package, an augmented plane-wave all-electron program[54]. We employed the Perdew-Burke-Ernzerhof (PBE) generalized gradient approximation (GGA) for the exchange-correlation functional[55].

The basis set size was determined by RKmax = 7.5. Γ-centered 16 × 16 × 4 and 16 × 16 × 2 k-point meshes were used for the pristine and interlayer-expanded $NbSe_2$ models, respectively. Spin-orbit coupling was not included in the calculations. The crystal structure for unintercalated $NbSe_2$ (space group *P*6$_3$/*mmc*, *a* = *b* = 3.4446 Å, *c* = 12.5444 Å) was taken from ICSD-16304 (ICSD release 2025.1)[56] in Inorganic Crystal Structure Database (ICSD)[57]. For the interlayer-expanded model, the same crystal structure was used, but the *c*-axis was expanded to 23.1700 Å, without introducing intercalated molecules.

**Raman measurements**

The Raman spectra were acquired with a home-built setup following the concept of previous work[58]. The excitation laser with wavelength of 532 nm was guided to the sample with free-space wave guide and focused to the sample with NA ~ 0.8 objective. The reflected laser beam was corrected by the same objective, guided through a set of narrow-band notch filters (Coherent), and dispersed by a spectrometer equipped with a CCD detector (Teledyne Prinston Instruments). The sample was mounted on three-axes piezo stage (attocube) to adjust laser spot position and focus. The laser power and the acquisition time was 700 μW and 120 s, respectively.

**ARPES measurements**

Angle-resolved photoemission spectroscopy (ARPES) was performed with a Scienta Omicron DA30-L electron analyzer and a VUV5000 helium discharge lamp with a photon energy of 21.2 eV (He Iα) at the University of Tokyo. The energy resolution was set to 15 meV. The Fermi level of the samples was referenced from the Fermi-edge spectra of a polycrystalline gold. The samples were cleaved *in situ* at 15 K. The measurements were performed while keeping the sample at 15 K under an ultrahigh vacuum better than $5 \times 10^{-10}$ Torr.

**Acknowledgements**     We thank F. Kagawa, Y. Ando, K. Matsuura and T. Nakashima for their discussions. We thank A. Nakano and C. Koyama for experimental supports. This work was supported by JSPS KAKENHI (Grants No. JP21H05235, JP21K13888, JP22K18317, JP23H04017, JP23H05431, JP23H05462, JP24H00417, JP24H01212, JP24K01285, JP24K01293, JP24H01652, JP24H01644, JP24K17006 JP25H02030, JP25H02126, JP25H02141), JST PRESTO (Grant No. JPMJPR25H9, JPMJPR20L5, JPMJPR24H8), JST FOREST (Grant No. JPMJFR2038, JPMJFR221V, JPMJFR2362), JST CREST (Grant No. JPMJCR23O3), the Mitsubishi Foundation, the Sumitomo Foundation, the Tanaka Kikinzoku Memorial Foundation, Asahi glass foundation, and the Murata Science and Education Foundation. The synchrotron radiation experiments were performed at SPring-8 with the approval of the Japan Synchrotron Radiation Research Institute (JASRI) (Proposal No. 2025A1505 and 2025A1998)

**Author contributions**  H.M., M.Su. and N.K. jointly conceived project. M.Su. synthesized the chiral ionic liquids. T.U. and N.K. grew the single crystals of $NbSe_2$. T.U. performed electrochemical experiment with support from H.M. and M.Su. T.U. and H.M. performed transport experiments. T.U., H.M., S.A., S.K., Y.N. and T.A. performed XRD measurements. T.U. and T.I. performed first-principles calculations. T.U., H.M., Y.Z. and T.M. performed Raman measurement. F.K., K.M., M.Sa. and K.I. performed ARPES measurement. N.K. organized the project. H.M. and N.K. wrote the draft with support from T.U., S.A., S.K. and T.A. All the authors discussed the results and commented on the manuscript.

**Competing interests**  The authors declare no competing interests.

**Data availability**  All the data presented in figures are available at UTokyo Repository: https://...

**Extended Figures**

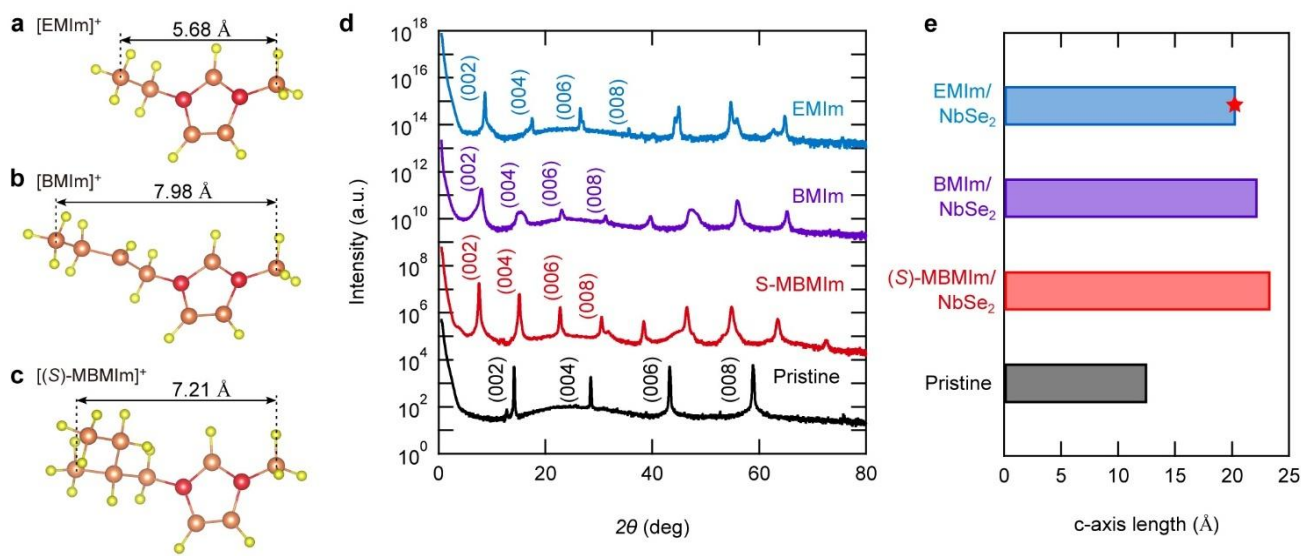

**Extended Data Fig. 1 | X-ray diffractions and $c$-axis lengths of $NbSe_2$ with several molecular-cation intercalants**

**a-c,** Schematic drawing of the molecular structures of [EMIm]$^+$ (1-Ethyl-2,3-dimethyl-imidazolium) (**a**), [BMIm]$^+$ (1-Butyl-2,3-dimethyl-imidazolium) (**b**), and [(*S*)-MBMIm]$^+$ ((*S*)-1- (2-Methylbutyl)-3-methylimidazolium ) (**c**). The lengths shown in the figures are defined as the distance between the end carbon atoms. **d,** Room-temperature X-ray diffraction (XRD) patterns of pristine (unintercalated) $NbSe_2$ and $NbSe_2$ intercalants of [(*S*)-MBMIm]$^+$, [BMIm]$^+$, and [EMIm]$^+$. **e,** Comparison of $c$-axis lengths estimated from XRD. The red star shows the $c$-axis length of previous reported EMIm/$NbSe_2$[21], which is consistent with our samples.

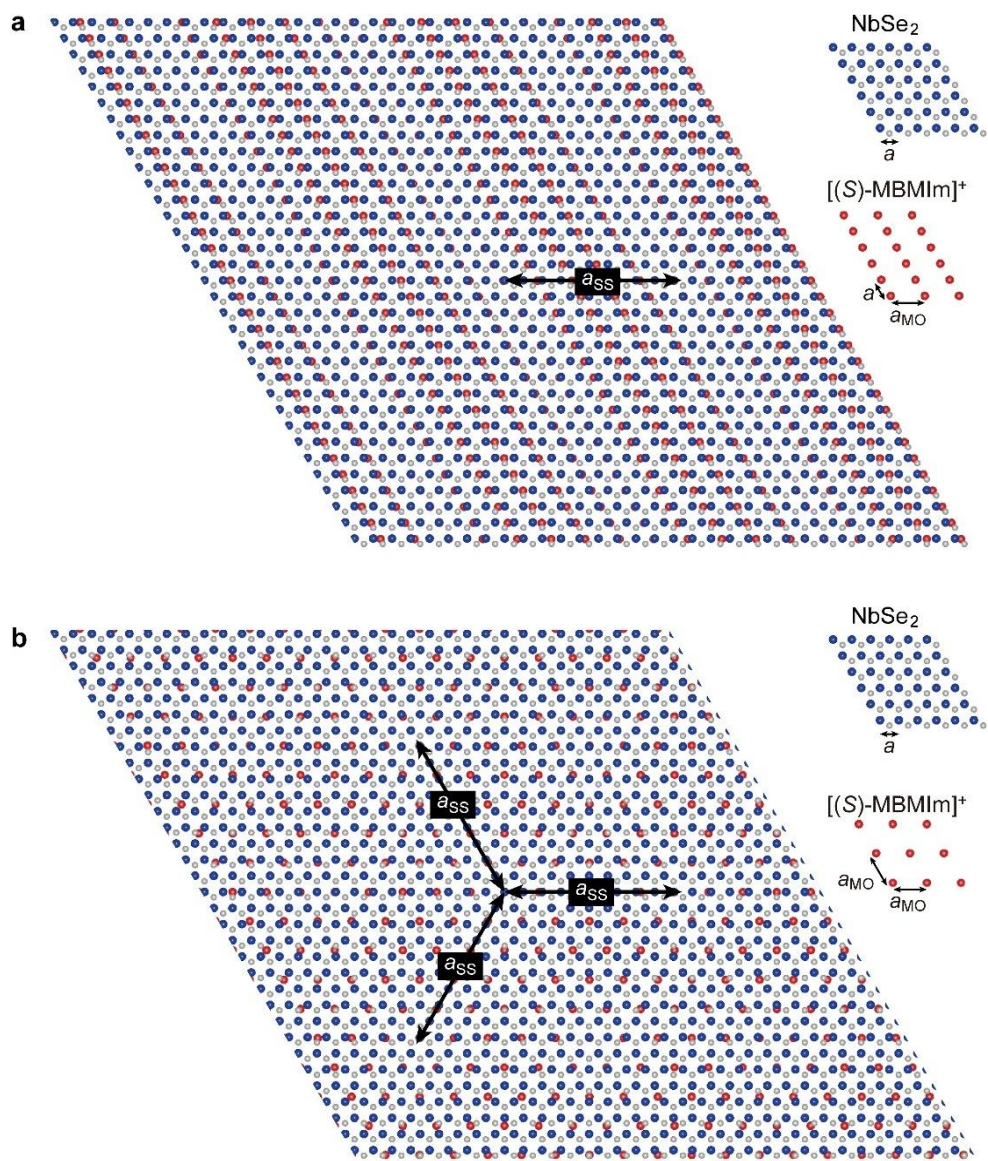

**Extended Data Fig. 2 | Moiré superlattices formed by stacking NbSe$_2$ and ordered organic molecules.**
**a,b,** Real-space moiré patterns simulated by stacking a single layer of NbSe$_2$ with a layer of organic molecules ordered in a single-$q$ configuration (**a**) and a triple-$q$ configuration (**b**). Nb and Se atoms in NbSe$_2$ are represented by blue and grey circles, respectively, while each organic molecule is depicted schematically as a single red circle. The lattice constants of NbSe$_2$, the ordered molecular layer, and the long-period superlattice are taken from experiment as $a$ = 3.46 Å, $a_{MO}$ = 6.29 Å, and $a_{SS}$ = 32.7 Å, respectively.

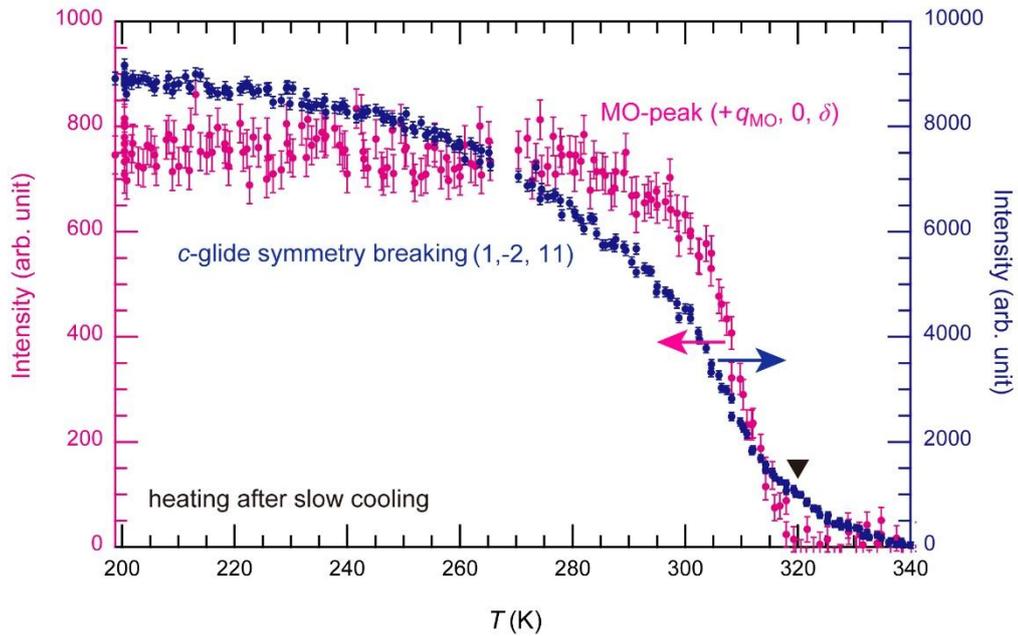

**Extended Data Fig. 3 | Temperature dependence of diffraction peaks in (*S*)-MBMIm/NbSe$_2$**

Temperature dependence of the diffraction intensities of the MO peak (+$q_{MO}$, 0, 1.1) and the (1,–2,11) peak during heating after slow cooling. The (1,–2,11) reflection, which is symmetry-equivalent to the (1,1,11) reflection by the $C_3$ rotational symmetry, belongs to the (*h*,*h*,*l*) (*l* is odd) reflections that are forbidden by the *c*-glide symmetry; therefore, its disappearance at high temperature corresponds to the restoration of the *c*-glide symmetry. Both peaks disappear simultaneously at the melting temperature $T_m$ as indicated by the black triangle, demonstrating that the breaking of the *c*-glide symmetry is intrinsically associated to the formation of the CSS phase.

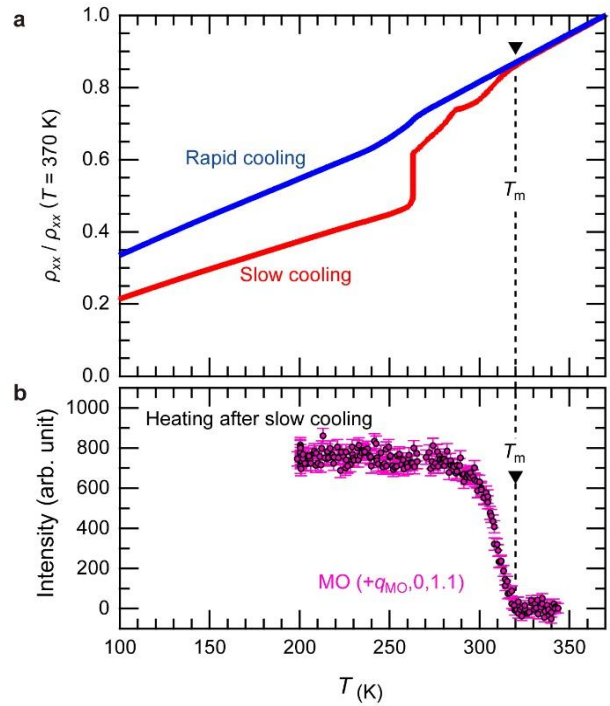

**Extended Data Fig. 4 | Correspondence between the CSS transition and resistivity anomaly in (*S*)-MBMIm/NbSe₂.**
**a,** Temperature dependence of the normalized resistivity of (*S*)-MBMIm/NbSe$_2$ measured under slow cooling (2 K min$^{-1}$, red) and rapid cooling (11 K min$^{-1}$, blue). The gray curve represents $d\rho_{xx}/dT$ under slow cooling. The onset of deviation between the two cooling curves, corresponding to the upper bound of the transition temperature $T_m$ = 320 K. **b,** Temperature dependence of the diffraction intensities of the molecular-ordering (MO) peaks measured on the same sample as in **a**. The disappearance of the MO peaks upon heating defines the melting temperature $T_m$, which closely matches the resistivity anomaly.

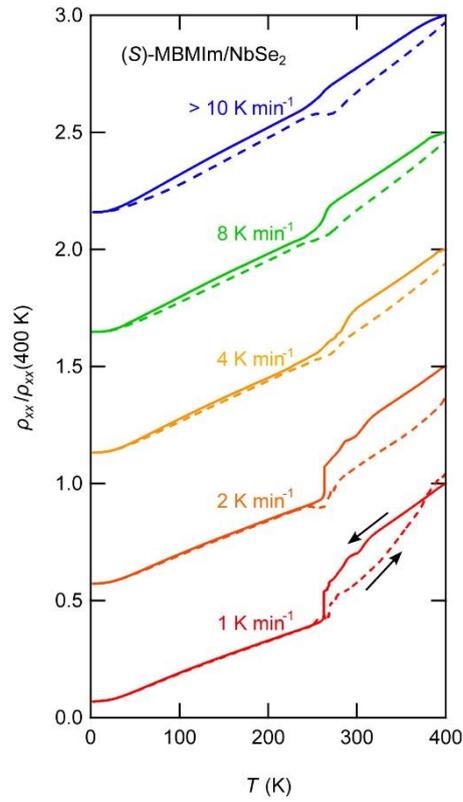

**Extended Data Fig. 5 | Temperature dependence of resistivity with varying sweep rate of temperature.**
Temperature dependence of resistance of (*S*)-MBMIm/NbSe$_2$ measured under various sweep rates. Solid lines denote the cooling process, and dashed lines denote the heating process, as indicated by the arrows.